\documentclass[12pt]{article}
\usepackage{amsmath,amssymb}
\usepackage{graphicx}

\parskip=\medskipamount

\title{Unfamiliar Aspects of B\"acklund Transformations and an Associated Degasperis-Procesi Equation}
\author{Alexander G. Rasin  \\
Department of Mathematics,\\ 
Ariel University, Ariel 40700, Israel \\
{E-mail: rasin@ariel.ac.il}\and  Jeremy Schiff \\
Department of Mathematics,\\
Bar-Ilan University, Ramat Gan, 52900, Israel \\
{E-mail: schiff@math.biu.ac.il}}

\begin{document}
\maketitle
\begin{abstract}
  We summarize the results of our recent work on B\"acklund transformations (BTs), particularly focusing on
  the relationship of BTs and infinitesimal symmetries. We present a BT for an associated Degasperis-Procesi
  (aDP) equation and its superposition principle, and investigate the solutions generated by
  application of this BT. Following our general methodology, we use the superposition principle
  of the BT to generate the infinitesimal symmetries of the aDP equation. 
\end{abstract}

\section{Introduction: B\"acklund Transformations and Symmetries}
B\"acklund transformations (BTs) are a characteristic feature of integrable differential equations, which are
typically regarded as curious, but not of fundamental importance. A prototypical example of a BT is that 
of the potential KdV (pKdV) equation: If $u$ satisfies the pKdV equation 
$$  u_t = \frac14 u_{xxx} + \frac32 u_x^2     $$
then so does $u_\alpha = u+v_\alpha$, where $v_\alpha$ is a solution of the system
\begin{eqnarray*}
    v_{\alpha,x}   &=& \alpha - 2 u_x  - v_\alpha^2 \ ,  \\
    v_{\alpha,t}   &=& u_{xx} v_\alpha - \frac12 u_{xxx} + (\alpha+u_x)(\alpha-2u_x -v_\alpha^2)   \ ,
\end{eqnarray*}
and $\alpha$ is a parameter.
Note that the equations of this system are {\em ordinary} differential equations for $v_\alpha$.
(It is not immediately obvious that the two equations are consistent --- but the consistency condition
turns out to be that $u_x$ satisfies the KdV equation, which is the case if  $u$ satisfies pKdV.) 
The  general solution of the first equation will depend on an arbitrary function of $t$,
and the second equation in the system will determine this up to an arbitrary constant. Thus $v_\alpha$
in fact depends on a second ``hidden'' parameter, as well as the parameter $\alpha$. In this sense
the BT, for a fixed value of the parameter $\alpha$, generates a $1$-parameter family of solutions $u_\alpha$
of pKdV for a given starting solution $u$.

A fundamental result on BTs is the {\em Bianchi permutability theorem}, that states that BTs commute. The
solutions $u_{\alpha,\beta}$ generated from the starting solution $u$ and first applying the BT with parameter $\alpha$
and then the BT with parameter $\beta$ coincide with the solutions $u_{\beta,\alpha}$ obtained by applying the BTs
in the reverse order. Furthermore we have the {\em algebraic superposition principle} 
  $$  u_{\alpha,\beta} = u_{\beta,\alpha} = u + \frac{\beta-\alpha}{u_\beta-u_\alpha}\ .  $$
In \cite{rs1} we realized the following simple fact: In the fraction in the last expression, if we let $\beta$ tend to  
$\alpha$, the numerator of the fraction will tend to zero, but the denominator need not tend to zero, because of the
dependence on different values of the hidden parameters in $u_\alpha$ and $u_\beta$. Thus
\begin{equation}
  Q(\alpha) = \frac1{u_{\alpha}^{(1)} - u_{\alpha}^{(2)}} \label{KdVQeq} \end{equation}
is an infinitesimal symmetry of pKdV.
Here $u_\alpha^{(1)}$ and $u_\alpha^{(2)}$ denote $2$ {\em distinct} solutions obtained from $u$ by application of the BT with
parameter $\alpha$.   Recall \cite{Ol0} that $\eta$ is an infinitesimal symmetry of pKdV
if both $u$ and $u+\epsilon \eta$ solve pKdV to first order in $\epsilon$, i.e. if
\begin{equation}  \eta_t = \frac14 \eta_{xxx} + 3 u_x \eta_x\ .     \label{symm} \end{equation}
It is straightforward to verify directly that $\eta = Q(\alpha)$ given in (\ref{KdVQeq})
satisfies (\ref{symm}). A general solution $\eta$ of (\ref{symm}) is called a {\em local} symmetry if it is
a function of the coordinates $x,t$,  the function $u$ and a finite number of derivatives of $u$. 
The symmetry of $Q(\alpha)$ is {\em nonlocal}; however we showed in \cite{rs1} that by expanding $Q(\alpha)$ in
an asymptotic series for large $\alpha$ we recover the standard infinite hierarchy of local symmetries of pKdV.
Furthermore we derived the standard recursion relation between the local symmetries \cite{Ol1} and
proved their commutativity. We also showed that the symmetry $Q(\alpha)$ includes known nonlocal symmetries
of pKdV. 

The fact that a BT can be used to derive local symmetries of an integrable equation was shown by Kumei in \cite{Kumei}
for the Sine-Gordon equation\footnote{We thank Peter Olver for bringing this reference to our attention.},
but this was thought to be an isolated case. In fact deriving local symmetries from BTs 
seems to be possible, using superposition principles, for many, if not all, integrable equations.
In \cite{rs2}  we
investigated the BT of the Camassa-Holm (CH) equation, following on the previous work \cite{Je1,rs1} for the associated CH
equation (an equation related to CH by a change of coordinates). Once again, the superposition principle for two
BTs was found to encode the infinitesimal symmetries. 
In \cite{rs3} we investigated the BT
of the Boussinesq equation. In this case the superposition principle that encodes the infinitesimal symmetries was
found to be a superposition for $3$ BTs. It seems natural to conjecture that this is related to the fact that 
the Lax pair for the Boussinesq equation is third order. 

A variety of other unfamiliar aspects of BTs emerged in these works. In the case of the CH equation \cite{rs2}
the BT acts not only on the dependent variable, but also on (one of the) independent variables.
In the case of the Boussinesq equation \cite{rs3} a single application of the BT produces not
only travelling wave solutions, but also what we called ``merging soliton''
solutions, in which two solitary waves merge into a single solitary wave.  Using the superposition principle we
found superpositions of merging solitons with solitons (thus, for example, giving a solution in which three solitary
waves merge to two), but superpositions of merging solitons with themselves seemed to always produce finite time
singularities (which in certain cases disappeared again at a later, but finite, time). 
The fact that BTs encode infinitesimal symmetries renews interest in BTs as a solution-generating technique. However, 
we note that we have still not succeeded to use BT techniques to generate multipeakon solutions of the CH equation. 


The aim of this paper is to show another example of many of these interesting features of BTs. 
In the next section we introduce and explain the siginificance of an equation we called 
the associated Degasperis-Procesi (aDP) equation. 
In Section 3 we give the BT and the superposition principle. Sections 4 and 5 are
devoted, respectively,  to solutions generated by the  application of $1$ and $2$ BTs. A single application of the
BT gives rise to 
two kinds of solitons, as well as a number of types of singular solutions, and also ``mergers'' in which a singular soliton
absorbs (either kind of) nonsingular soliton 
(mergers of nonsingular solitons do not seem to be allowed). By application
of a second BT these solutions can be superposed, in this case (it seems) without any new singularity emerging. 
In section 6 we use the superposition principle of 3 BTs
to derive the infinitesimal symmetries of the aDP equation; the first nontrivial symmetry is the Kaup-Kupershmidt (KK)
equation (this being in accord with the results of Degasperis, Holm and Hone \cite{add4}, who related aDP to the KK hierarchy).
In section 7 we conclude.
In an appendix, we  briefly give a direct derivation of the travelling wave solutions of aDP.  

\section{The associated DP equation}

The Degasperis-Procesi equation \cite{DP} is the equation
\begin{equation}
  u_T - u_{XXT} + 4uu_X - uu_{XXX} - 3u_Xu_{XX}  = 0 \ . \label{DP}
\end{equation}     
This equation is the case $b=3$ of the so-called ``$b$-family'' of equations 
\begin{equation}
  m_T + um_X + bmu_X = 0 \ , \qquad m = u-u_{XX} \ .  \label{beq}
\end{equation}   
In \cite{DGGH1} this family of equations (except for the case $b=-1$)  
was shown to arise from a shallow water equation via a Kodama transformation.
For an extensive study of the $b$-family see \cite{HS1,HS2}. 
For $b>1$ there are stable peakon solutions  $u(X,T)=ce^{-|X-cT|}$, which can be superposed to give 
more general solutions of the form
$$  u(X,T) = \sum_{i=1}^N  p_i(T) e^{-|X-q_i(T)|} $$
provided the positions $q_i(T)$ and speeds $p_i(T)$ satisfy the equations of motion of a
certain finite dimensional Hamiltonian system. In \cite{HS2} it is suggested that
$b=3$ is a possible bifurcation point of the $b$-family. Furthermore, it is known that there 
are only two integrable cases of the $b$-family, the case $b=2$,
for which it reduces to the CH equation, and the case $b=3$ of DP.

Integrability in the case $b=3$ was established in \cite{add4}, where a Lax pair was given, and
a bihamiltonian structure proposed, for an equation related to DP by a change of coordinates.
The bihamiltonian structure was validated in \cite{HW2}, and further evidence presented
singling out the cases $b=2$ and $b=3$.  A version of the DP equation after the change of coordinates
is what we call  here the associated DP (aDP) equation, so we present its derivation in full detail. 
Equation  (\ref{beq}) can be written in the form 
$$    (m^{1/b})_T  +  (m^{1/b} u)_X  = 0 \ .     $$
Writing $p = m^{1/b}$, this implies we can define new coordinates $x,t$ via
$$  dx = p dX - p u dT \ , \qquad  dt = dT\ .    $$
Moving to the new coordinates we find that $u$ can be eliminated via 
$$   u =  p^b - p_{xt} + \frac{p_xp_t}{p}  $$ 
and there remains a single equation for $p$, that can be written in either of the forms
$$  \left(\frac1{p}\right)_t   +   \left( p_{xt}  -  \frac{p_xp_t}{p}  - p^b   \right)_x = 0    $$ 
or 
$$  \left(\frac{p_{xx}}{p} - \frac{p_x^2}{2p^2} + \frac1{2p^2}  \right)_t  + \left(\frac{bp^{b-1}}{b-1} \right)_x   = 0\ .   $$
The second form suggests introducing a new unknown $f(x,t)$ defined by  $p = f_t^{1/(b-1)}$. The equation for $f$ is then
$$  f_{xxt} - \frac{(2b-3)f_{xt}^2}{2(b-1)f_t}   +   c_1 f_xf_t + c_2 f_t^{\frac{b-3}{b-1}}  = 0   $$ 
where $c_1,c_2$ are constants, which can be chosen freely by rescalings of $f$, $x$ and $t$.
Making a convenient choice we arrive, in the case $b=3$, to what we call for the purposes of this paper
the {\em associated DP} (aDP) equation: 
\begin{equation}   
  f_{xxt} - \frac34 \frac{f_{xt}^2}{f_t} +  3( 1 - f_x f_t )  = 0  \ .    \label{myDP}
\end{equation}
The reason we do not study DP directly, but rather work with aDP,  is that it turns out that formulas for the BT are
simpler when
expressed in terms of $f$ than in terms of the other fields, {\it viz.} $u,m,p$.
Specifically, in \cite{add4} it was established that after the change of coordinates we have described, the DP
equation can be considered as a negative flow in the Kaup-Kupershmidt (KK) hierarchy. The KK variable in
the notation of \cite{add4} is the field $V$, which in our notation can be identified, modulo rescaling,  with $f_x$.
Thus it will come as no surprise that in section 6 we find the first nontrivial symmetry of (\ref{myDP}) is the
potential KK equation. The connection between the DP and KK equations was further explored in the recent
paper \cite{Ol2}. 

One motivation for the study of the BT of aDP (\ref{myDP}) is to advance the state of knowledge
of solutions of DP (\ref{DP}). A large number of solutions of DP are already known. Multipeakon  solutions
are discussed in \cite{add4} and \cite{LS1}. Travelling wave solutions are discussed in 
\cite{VP1} and \cite{Le1}, the former including the description of ``loop-like''  solitary
waves and periodic solutions. In the formidable papers \cite{Ma1,Ma2} Matsuno uses the relation of the DP with the
KK hierarchy, the fact that the KK hierarchy is a reduction of the CKP hierarchy, and known results about CKP
tau functions to derive multisoliton solutions of DP. A similar approach appears in works of Feng {\em et al.}
\cite{FMO1,FMO2}, and the paper \cite{SS1} uses an extension of Matsuno's results to describe superpositions
of loop solitons and mixed superpositions of solitons and loop solitons. In the current paper, however,  we
focus on application of the BT of aDP and defer comparison with existing results on DP for a later work. 


A second  motivation for the study of the BT of aDP (\ref{myDP}) is the link to the KK equation.
As far as we know, a BT for the KK equation is not known, and we believe it should be possible 
to extend the BT given here for aDP to KK.  Furthermore, KK is a piece of larger picture: the KK and the Sawada-Kotera (SK) equations
are both reductions of a general Lax equation describing the evolution of a
3rd order Lax operator \cite{SAM1}. KK and SK are related, both being generated 
by Miura maps from a common ``modified'' equation \cite{FG1,FG2,FG3}. In the same way
as the DP equation is related to KK, the Novikov equation \cite{No1} is related to SK, as 
demonstrated in \cite{HW1}, by a change of coordinates similar to the one we have presented for DP. 
We note in passing that introducing the function $f$ as described above
for the Novikov equation yields the extremely simple ``associated Novikov'' (aN) equation
\begin{equation}   
  f_{xxt} +  3( 1 - f_x f_t )  = 0  \ .    \label{myNov}
\end{equation}
We believe there is analog of our results for the aDP equation for the aN  equation.
But once again, we defer discussion of BTs for the full set of aDP, KK, aN and SK equations, and their interrelations,
and the BT for the general Lax system from which they are obtained as a reduction, to  a later work. 
Note that a BT for SK is known \cite{SAM1}. The interrelations of the
DP, KK, Novikov and SK were studied in \cite{Ol2}, with focus on Hamiltonian structures. 

\section{The BT and the Superposition Principle} 

A direct calculation verifies that the  aDP equation (\ref{myDP}) has a BT that can be written in either the form 
\begin{equation} 
  f \rightarrow f -  \frac{2\theta}{s^2 + 2s_x - 3f_x}
  \qquad {\rm or} \qquad  
  f \rightarrow f - 2 q
\end{equation} 
where $s$ satisfies
\begin{eqnarray}
  s_{xx}  &=& \theta - s^3 - 3ss_x   + 3f_x s + \frac32 f_{xx} \ ,  \label{sxx}\\
  s_t &=&  f_t + \frac1{\theta}\left(   \frac12 f_{xt}s_x - f_tss_x
  - f_t s^3 + f_{tx} s^2 + \left(\frac32+ \frac{3f_xf_t}{2}   - \frac{3f_{xt}^2}{8f_t} \right) s \right. \nonumber \\
  && \left. 
     -  \frac{3f_xf_{xt}}{4} + \frac{3f_{xt}}{4f_t}   +  \frac{f_{xt}^3}{16f_t^2}        \right) \ ,  \label{st}
\end{eqnarray}
and $q$  satisfies
\begin{eqnarray}
  q_{xx}  &=& \theta - q^3 + \frac34 \frac{q_x^2}{q} - 3 qq_x + 3q f_x \ ,  \label{qxx}\\
  q_t    &=& \frac12 f_t - \frac1{2\theta^2f_t} \left(  q^3f_t  + f_tqq_x  - q^2 f_{xt} + \frac{f_tq_x^2}{4q} -  \frac12 q_xf_{xt}
                  + \frac{qf_{xt}^2}{4f_t} - 3q   \right)^2  \ .  \label{qt}
\end{eqnarray}
The advantages of each of the formulations are clear: the action of the BT is simpler in the $q$ formulation,
but the equations satisfied by $q$ are more complicated than those satisfied by $s$
(see below, equation(\ref{conslaw}), for a simpler version of equation (\ref{st})).  The relationships between $q$ and $s$
are
$$   q =  \frac{\theta}{s^2 + 2s_x - 3f_x}   \qquad{\rm and} \qquad   s = \frac{q_x}{2q} + q\ .   $$
Note that the equations for $s$ (or for $q$) can be linearized via the substitution $s=\frac{\psi_x}{\psi} $ 
yielding 
\begin{eqnarray}
  \psi_{xxx} &=&  3 f_x \psi_x + \left(  \frac32 f_{xx} + \theta \right) \psi \ ,  \label{psixxx} \\
  \psi_t    &=& \frac1{\theta}\left(  f_t \psi_{xx} - \frac12 f_{tx} \psi_{x}  +
   \left( \frac18 \frac{f_{xt}^2}{f_t} - \frac32 f_x f_t \right) \psi 
       \right)
     \ .  \label{psit}
\end{eqnarray}
This is the Lax pair for the aDP equation (compare \cite{add4}). 
Note that equation (\ref{psixxx}) can also be written in the form
$$  \left( \psi \psi_{xx} - \frac12 \psi_x^2 - \frac32 f_x \psi^2 \right)_x   = \theta \psi^2   $$ 
so
$$   \psi \psi_{xx} - \frac12 \psi_x^2 - \frac32 f_x \psi^2     = \theta \int \psi^2\ dx  \ .   $$ 
and thus  
\begin{equation}
q
= \frac{\theta}{2 \frac{\psi_{xx}}{\psi} - \frac{\psi_x^2}{\psi^2} - 3f_x  }  
=  \frac{\psi^2}{2 \int \psi^2 \ dx}\ .   \label{qeq}
\end{equation}

We mention in passing that the BT can be used to find  conservation laws for the aDP equation. From 
(\ref{sxx})-(\ref{st}) it follows that
\begin{equation}
  s_t = \frac{1}{\theta} \left( f_t(s_x+s^2) - \frac12 f_{xt} s - \frac32 f_xf_t + \frac{f_{xt}^2}{8f_t}  \right)_x
\label{conslaw}\end{equation}  
which is a $\theta$ dependent conservation law for aDP. 
For large $|\theta|$, $s$ can be expanded in an asymptotic series
\begin{equation}
  s \sim  \sum_{i=0}^\infty  s_i(x,t) \theta^{(1-i)/3} \ , \qquad s_0(x,t)=1\ . \label{asymp} 
\end{equation} 
Substituting this series into (\ref{sxx}) gives a recursion relation for the coefficients $s_i(x,t)$ in terms
of $f$ and its $x$-derivatives. Once the coefficients are determined, substituting the series into (\ref{conslaw})
and expanding in powers of $\theta^{-1/3}$ gives an infinite sequence of conservation laws. Many of these
are trivial. The first nontrivial one (after a little cleaning up) is
$$  \left(  f_{xx}^2  +  4 f_x^3 \right)_t =
\left( 3f_tf_x^2 - 6f_x +  \frac{3f_xf_{xt}^2-18}{2f_t}  +   \frac{3f_{xt}^2}{2f_t^2}  -  \frac{f_{xt}^4}{16f_t^3}
\right)_x 
$$ 

We note the BT for aDP has an intriguing superficial similarity to the BT for the Boussinesq equation (equations (8)-(9) 
in \cite{rs3}). However the BT was actually discovered by trial and error searching for a BT for
DP that was superficially similar to the BT for CH given in \cite{rs2}. 

The BT has a superposition principle  which we present  in the $q$ formulation: 
If $q_1,q_2$ are (respectively) solutions of (\ref{qxx})-(\ref{qt})
with parameters $\theta_1,\theta_2$, then the solution obtained by applying the 2 BTs with parameters
$\theta_1$ and $\theta_2$, in either order, is
\begin{equation}
f_{12} = f  - 
\frac{ 2 (\theta_1+\theta_2)  \left( 2Z + (\theta_1-\theta_2)\left(\frac1{q_1}-\frac1{q_2} \right) \right) }
     { \left(  Z  + \left(\frac{\theta_1}{q_1}+\frac{\theta_2}{q_2}\right)  \right)^2 -  \frac{(\theta_1+\theta_2)^2}{q_1q_2}}
\label{superpos}
\end{equation}
where
$$  Z =  \left(\frac{q_{1x}}{2q_1} - \frac{q_{2x}}{2q_2} +  q_1-q_2 \right)^2  =  \left( s_1-s_2 \right)^2\ .   $$
This result was also found by trial and error, but once known, can easilly be  verified directly. 

\section{Solutions from a Single BT}

We apply the BT to the ``trivial'' solution 
$$   f =  \beta x + \frac{t}{\beta}  + \alpha $$
where $\alpha, \beta$ are constants with $\beta\not=0$.  The general solution of (\ref{psixxx})-(\ref{psit}) is 
$$ \psi =  C_1 e^{\lambda_1 x+\frac{t}{\beta\lambda_1}}
+ C_2 e^{\lambda_2 x+\frac{t}{\beta\lambda_2}}
+ C_3 e^{\lambda_3 x+\frac{t}{\beta\lambda_3}}  $$
where $\lambda_1,\lambda_2,\lambda_3$ are the $3$ roots of  $\lambda^3 = 3 \beta\lambda + \theta$, 
and $C_1,C_2,C_3$ are constants.  Using (\ref{qeq}) this gives us new solutions 
\begin{equation}
f =  \beta x + \frac{t}{\beta} -  \frac{2(\eta_1 +\eta_2+\eta_3)^2} 
{ \frac{\eta_1^2}{\lambda_1} + \frac{\eta_2^2}{\lambda_2} + \frac{\eta_3^2}{\lambda_3}
- \frac{4\eta_1\eta_2}{\lambda_3} - \frac{4\eta_2\eta_3}{\lambda_1} - \frac{4\eta_3\eta_1}{\lambda_2}
}  \label{1BT}
\end{equation}
where
$$\eta_i  = C_i e^{\lambda_i x+\frac{t}{\beta\lambda_i}} \ , \quad i=1,2,3 \ . $$ 

Analysis of these solutions proceeds in a similar manner as in \cite{rs3}. We restrict to the case $4\beta^3 > \theta^2$, in
which case all the roots $\lambda_1,\lambda_2,\lambda_3$ are real. If two of the constants $C_i$ vanish, the solution is trivial. 
If exactly one of the constants $C_i$ vanishes then the solution is a travelling wave, with speed $\frac{\lambda_i}{\beta\theta}$
(where $\lambda_i$ is the root of the same index as the coefficient $C_i$ that vanishes). 
There are two kinds of (right moving) soliton solution and two kinds of
(left moving) singular soliton solution; the  wave profiles are displayed in Figure 1.
One of the solitons has a regular soliton profile --- we call this a ``simple soliton''. 
The other one has ``dimples'', at which the profile drops to zero, before
and after the main mass of the soliton --- we call this a ``dimpled soliton''. The singular solitons all have a single dimple,
but differ in whether it is on the left or the right of the main soliton mass. See the appendix for a direct derivation of the
travelling wave solutions of aDP, and more useful formulae. 

\begin{figure}
      \centerline{
        \includegraphics[width=7cm]{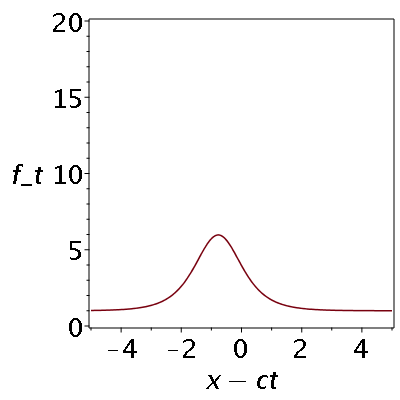}
        \includegraphics[width=7cm]{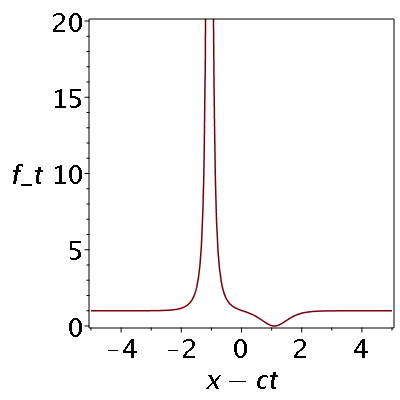} 
        }
      \centerline{
        \includegraphics[width=7cm]{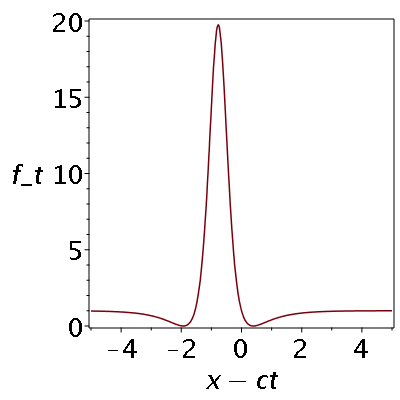}
        \includegraphics[width=7cm]{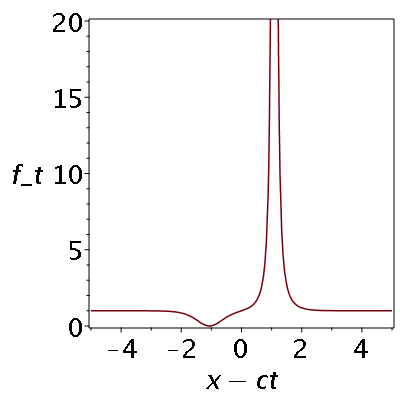} 
        }
      \caption{Profles of travelling wave solutions of aDP (\ref{myDP}). Plots are of $f_t$ as a function of $x$, with $f$ given
        by (\ref{1BT}). Parameters are $\beta=1$, $\theta=\frac12$,  so $\lambda_1\approx -1.64$, $\lambda_2\approx -0.17$,
        $\lambda_3 \approx 1.81$. 
       Top left: $C_1=1$, $C_2=1$, $C_3=0$, solution moves right, the ``simple soliton''. 
       Top right:   $C_1=1$, $C_2=0$, $C_3=1$, solution moves left. 
       Bottom left: $C_1=1$, $C_2=-1$, $C_3=0$, solution moves right, the ``dimpled soliton''. 
       Bottom right: $C_1=1$, $C_2=0$, $C_3=-1$, solution moves left. 
      }
\end{figure}

The solutions with all of the $C_i$ nonzero are not  travelling wave solutions.
They describe the absorption of one of the two types of soliton 
by one of the two types of singular soliton. When a simple soliton gets absorbed, by
either of the types  of singular soliton, the type of the singular soliton remains the
same. But when a dimpled soliton gets absorbed, then the type of the singular soliton
is switched. See Figures 2,3 for two examples.  

The dimpled soliton solutions are to some extent an artefact of the fact we are plotting $f_t$ as the wave height.
In terms of the variable $p$ satisfying   $f_t=p^2$ the dimpled solitons appear as antisolitons. But then there is an
asymmetry between solitons and antisolitons: whereas solitons can have arbitrarilly small elevation over the background level,
antisolitons have a minimum depth (corresponding to $f_t$ dropping from $\frac1\beta$ to $0$). 

\begin{figure}
      \centerline{
        \includegraphics[width=4cm]{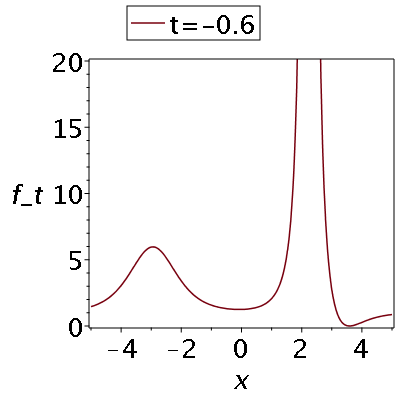} 
        \includegraphics[width=4cm]{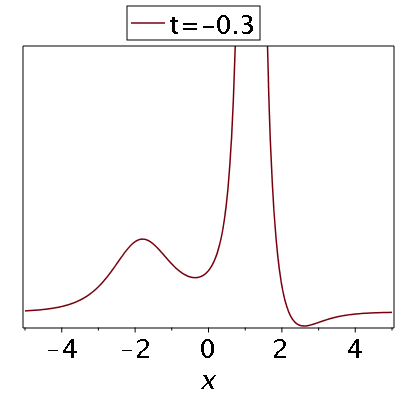} 
        \includegraphics[width=4cm]{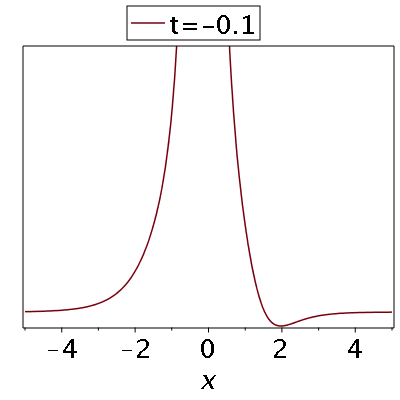} 
        }
      \centerline{
        \includegraphics[width=4cm]{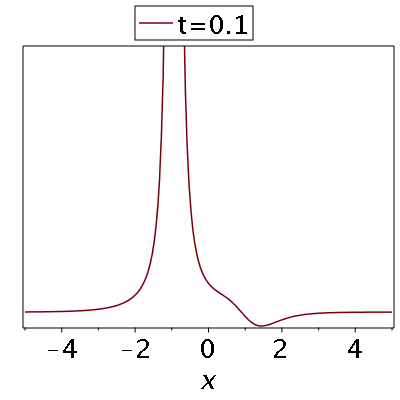}
        \includegraphics[width=4cm]{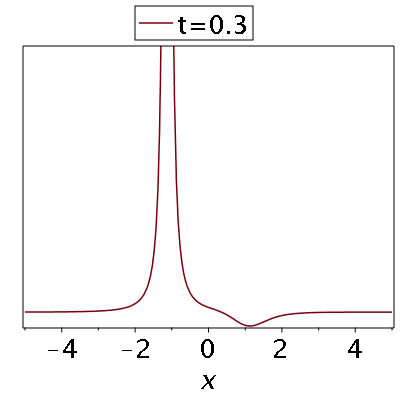}
        \includegraphics[width=4cm]{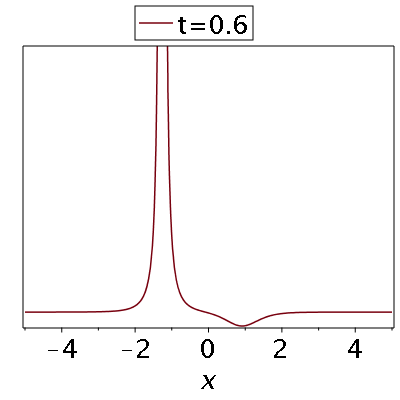} 
        }
      \caption{Absorption of simple soliton by singular soliton.  Parameter values as in Figure 1, but 
        $C_1=C_2=C_3=1$.  Times shown are    $t=-0.6,-0.3,-0.1,0.1,0.3,0.6$. 
      Note the type of the singular soliton
      is the same before and after absorption of the soliton --- the dimple remains on the right. }
\end{figure}


\begin{figure}
      \centerline{
        \includegraphics[width=4cm]{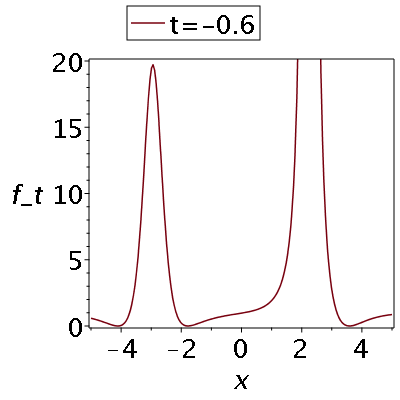}
        \includegraphics[width=4cm]{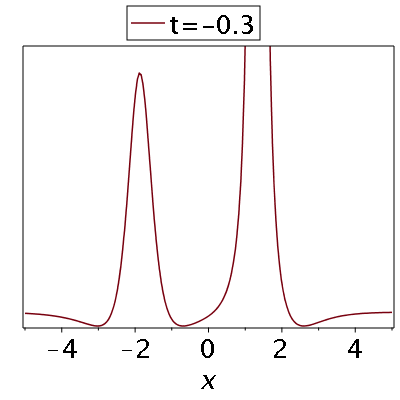}
        \includegraphics[width=4cm]{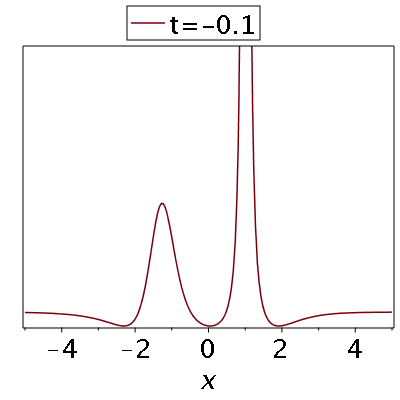} 
        }
      \centerline{
        \includegraphics[width=4cm]{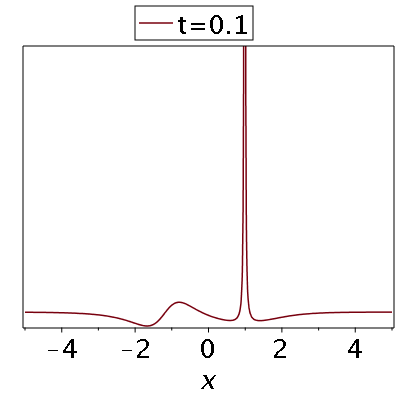} 
        \includegraphics[width=4cm]{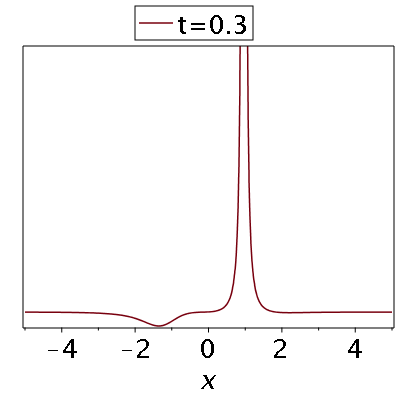} 
        \includegraphics[width=4cm]{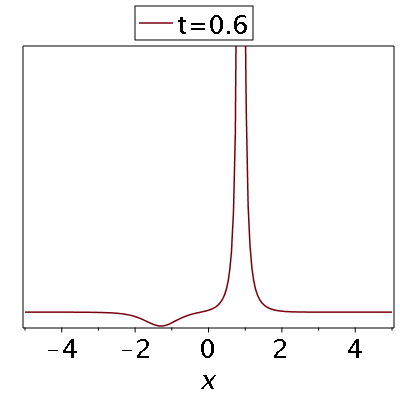} 
        }
      \caption{Absorption of dimpled soliton by singular soliton. Parameter values as in Figure 1, but 
        $C_1=-1$, $C_2=C_3=1$.  Times shown are   $t=-0.6,-0.3,-0.1,0.1,0.3,0.6$. 
       Note the type of the singular soliton
       changes as a result of absorption of the soliton --- the dimple moves from the right to the left.
       Note also that at a certain time a pair of dimples disappears, by colliding with a singularity.} 
\end{figure}


\section{Solutions from Double BTs}
Experiments with the superposition formula (\ref{superpos}) show that a wide range of 
different kinds of solution can be obtained using double BTs.  It is {\em not} the case that the physical content
of a 2 BT solution is the ``sum'' of the physical content of the solutions obtained by applying the 2 BTs individually ---
the superposition is highly nonlinear.

It is possible to superpose simple solitons with simple solitons,  dimpled solitons with dimpled solitons, and also
simple with dimpled. The process by which a fast dimpled soliton passes a slow simple soliton is very straightforward: the dimpled
soliton absorbs the simple soliton on its right, and later emits it on its left, as shown in Figure 4. The process by which
a fast simple soliton passes a slow dimpled soliton is much more complicated and involves 3 stages (see Figure 5):
(1) As the simple soliton approaches the dimpled soliton it starts to absorb it. Its height grows rapidly, while the height
of the hump between the two dimples drops dramatically. (But, it should be emphasized, the dimples never disappear at any
stage of the process, even though they get closer together, and the  height of the hump between them becomes very low.)
(2) When the hump between the two dimples has become sufficiently low, a new hump starts to grow on their right, while
the hump on their left (the enlarged incoming simple soliton) decays. This stage is extremely rapid. (3) Once the hump on the left
of the dimples has totally decayed, and the hump on their right is fully grown, the latter hump starts to decay again, while the hump
between the dimples regrows, to assume its original size. 

\begin{figure}
      \centerline{
        \includegraphics[width=4.6cm]{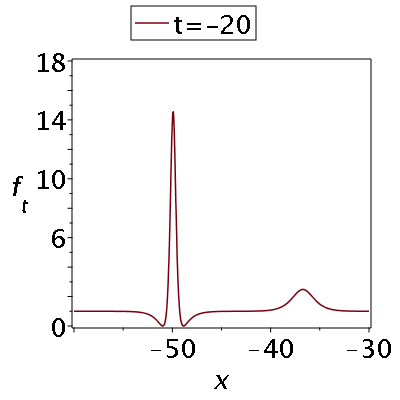} 
        \includegraphics[width=4.6cm]{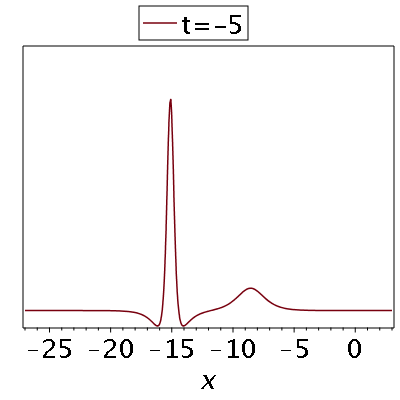} 
        \includegraphics[width=4.6cm]{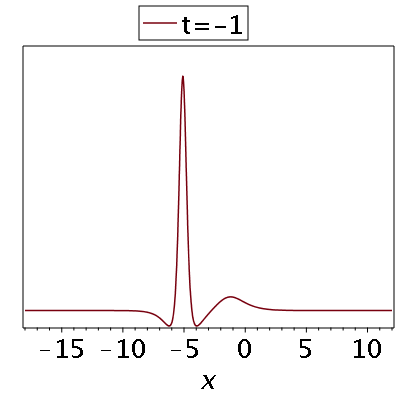} 
        }
      \centerline{
        \includegraphics[width=4.6cm]{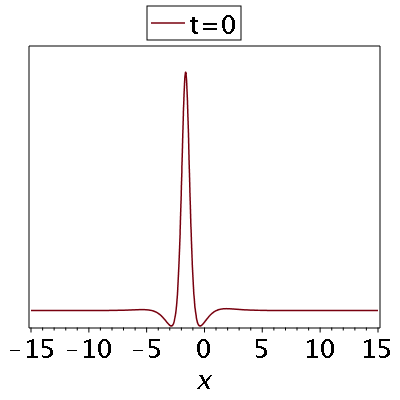} 
        \includegraphics[width=4.6cm]{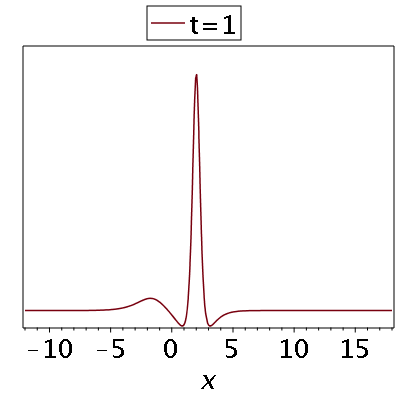} 
        \includegraphics[width=4.6cm]{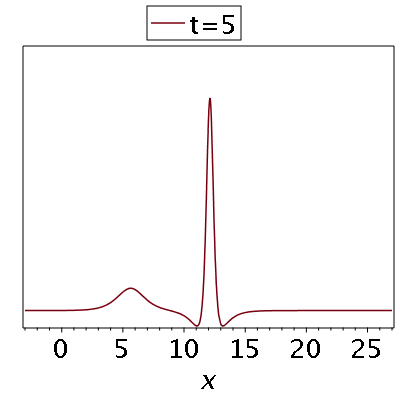} 
        }
      \caption{Superposition of fast dimpled soliton and slow simple soliton. The dimpled soliton absorbs the simple soliton
        and emits it on the other side. Parameters are $\beta=1$;  $\theta=0.8$, $C_1=C_2=1$, $C_3=0$ for $q_1$;
        $\theta=1.0$, $C_1=C_2=1$, $C_3=0$ for $q_2$.}
\end{figure}

\begin{figure}
      \centerline{
        \includegraphics[width=4.2cm]{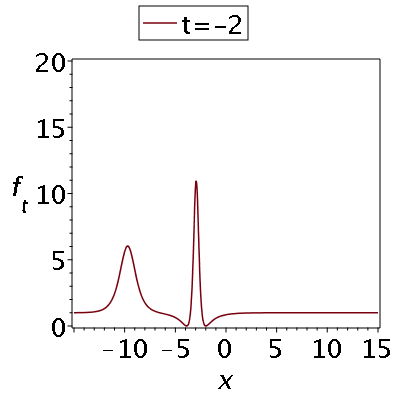} 
        \includegraphics[width=4.2cm]{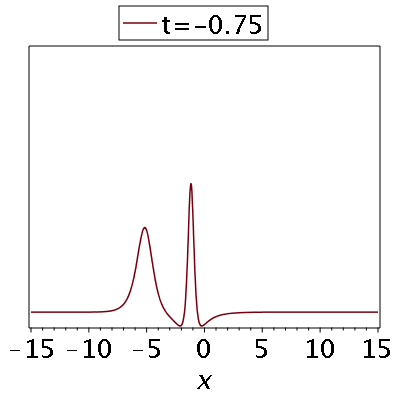} 
        \includegraphics[width=4.2cm]{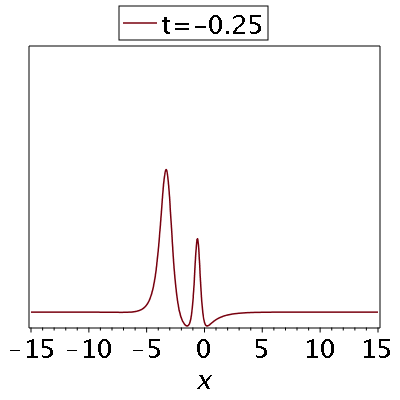} 
        \includegraphics[width=4.2cm]{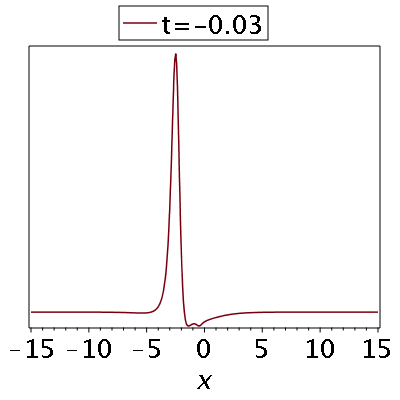} 
        }
      \centerline{
        \includegraphics[width=4.2cm]{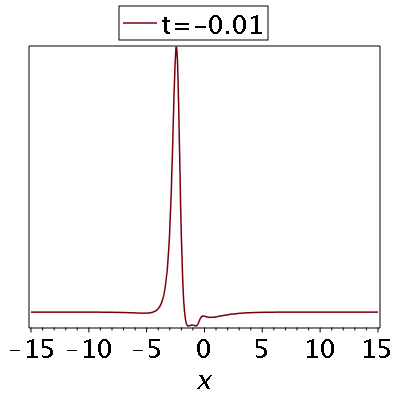} 
        \includegraphics[width=4.2cm]{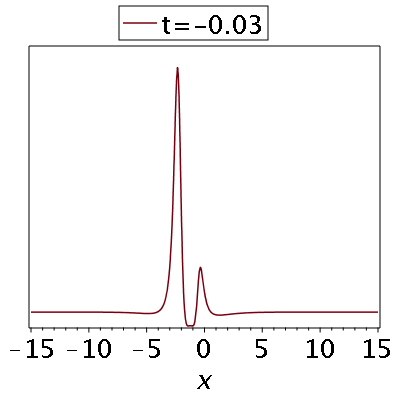} 
        \includegraphics[width=4.2cm]{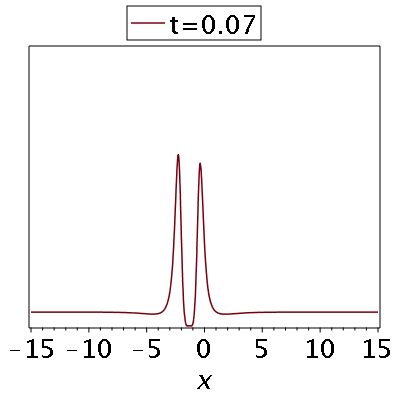} 
        \includegraphics[width=4.2cm]{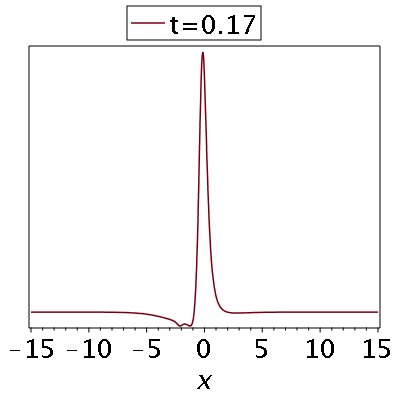}  
        }
      \centerline{
        \includegraphics[width=4.2cm]{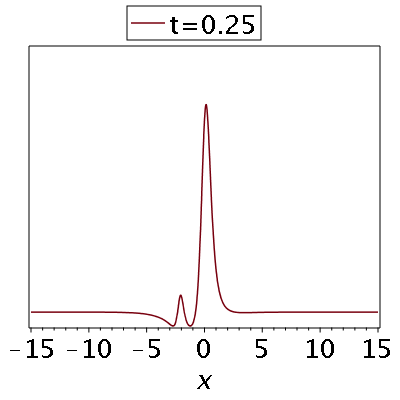} 
        \includegraphics[width=4.2cm]{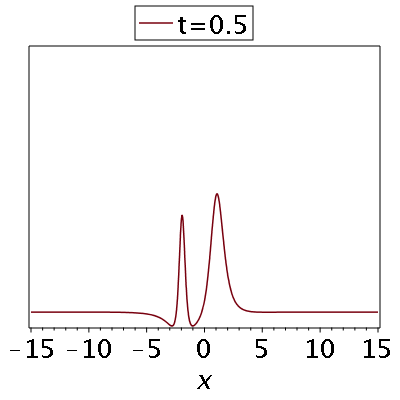}
        \includegraphics[width=4.2cm]{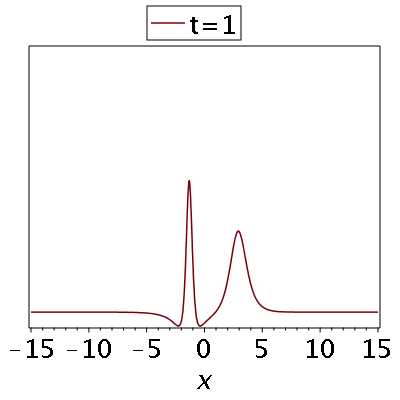}
        \includegraphics[width=4.2cm]{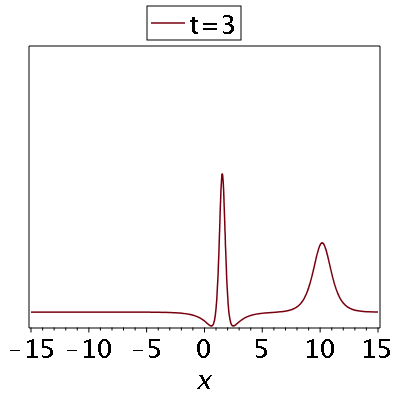}  
      }
      \caption{Superposition of fast simple soliton and slow dimpled soliton.  
        Parameters are $\beta=1$;  $\theta=\frac12$, $C_1=1$, $C_2=-1$, $C_3=0$ for $q_1$;
        $\theta=\frac43$, $C_1=1$, $C_2=-1$, $C_3=0$ for $q_2$.} First row: reduction of the central
      hump. Second row: emergence of the humo on the right and annihilation of the hump on
      the left. Third row: reemergence of the central hump. 
\end{figure}

We give just one further example of a superposition. Recall the merger solution illustrated in Figure 2, in
which a simple soliton is absorbed by a singular soliton. There are also superpositions with an initial state
involving a simple soliton and a singular soliton (of the same type as the example in Figure 2). In the
superposition these do not merge, but after collision, the simple soliton emerges on the other side of the singular
soliton as a dimpled soliton, and also the type of the singular soliton is changed. 
See Figure 6. As part of this process, a pair of new dimples is formed, at a point of singularity. 

\begin{figure}
      \centerline{
        \includegraphics[width=4.2cm]{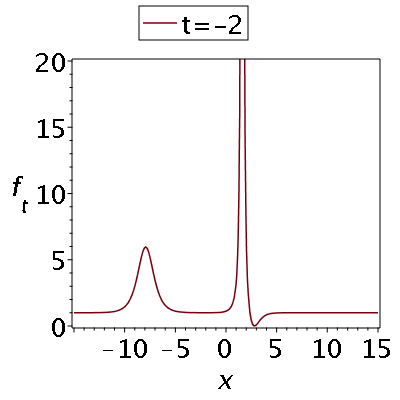} 
        \includegraphics[width=4.2cm]{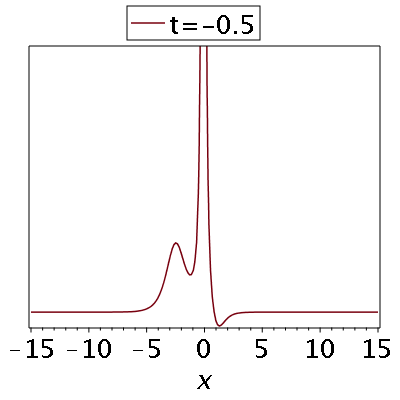} 
        \includegraphics[width=4.2cm]{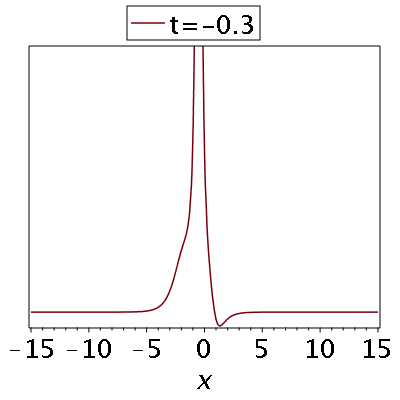} 
        \includegraphics[width=4.2cm]{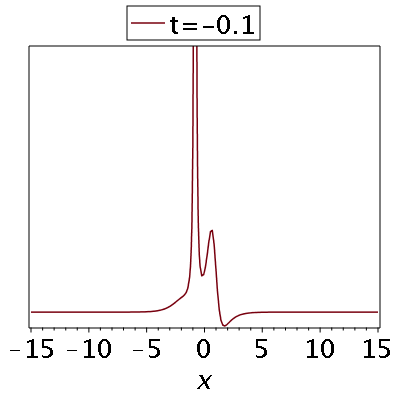} 
        }
      \centerline{
        \includegraphics[width=4.2cm]{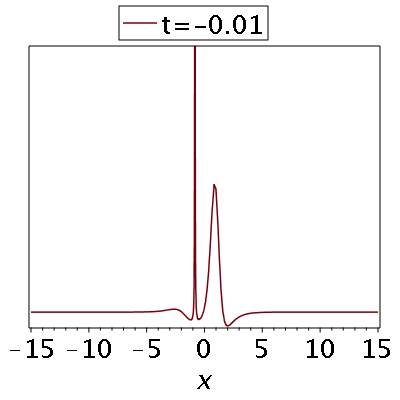} 
        \includegraphics[width=4.2cm]{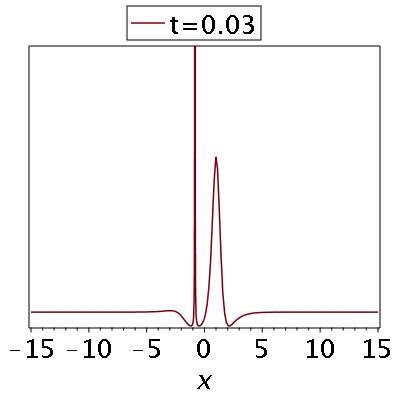} 
        \includegraphics[width=4.2cm]{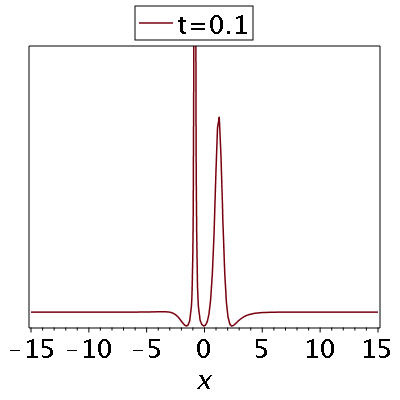} 
        \includegraphics[width=4.2cm]{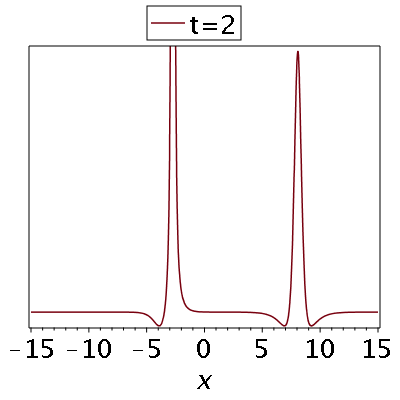}  
        }
      \caption{Superposition describing a simple soliton colliding with a singular soliton,
          and emerging as a dimpled soliton. 
        Parameters are $\beta=1$;  $\theta=\frac12$, $C_1=1$, $C_2=-1$, $C_3=0$ for $q_1$;
        $\theta=\frac43$, $C_1=0$, $C_2=1$, $C_3=1$ for $q_2$.}
\end{figure}

\section{Symmetries}

It is possible to verify that the double BT described by (\ref{superpos}) does not include the identity transformation,
for any choice of $\theta_1,\theta_2$ or the hidden parameters in $q_1,q_2$. Thus we are led to look at triple BTs.
We have not, as of yet, succeeded to find algebraic formulas for triple superpositions, along the lines of those
found for the Boussinesq equation in \cite{rs3}, though it is natural to conjecture there should be some relation
with the discrete CKP equation, as described in \cite{FN2}. However, it is possible to compute the effect of a triple BT
in the limit that the 3 parameters $\theta_1,\theta_2,\theta_3$ tend to a common limit $\theta$. In this limit the
triple BT gives  an infinitesimal symmetry 
\begin{equation}
  Q(\theta) =   \frac{s^{(1)}-s^{(2)}} { (s^{(1)}-s^{(2)})s^{(3)}_x + (s^{(2)}-s^{(3)})s^{(1)}_x + (s^{(3)}-s^{(1)})s^{(2)}_x
   - (s^{(1)}-s^{(2)})(s^{(2)}-s^{(3)})(s^{(3)}-s^{(1)})
  }  \label{gensym}
\end{equation} 
where $s^{(1)},s^{(2)},s^{(3)}$ are three distinct solutions of (\ref{sxx})-(\ref{st}). Remarkably, this is exactly the form of 
generating function for symmetries found for the Boussinesq equation in \cite{rs3}, equation (32). Even though the calculation
to arrive at the result (\ref{gensym}) is enormous, once the result has been found it is straightforward to check directly that
$\eta = Q(\theta)$ solves the linearized aDP equation
\begin{equation}   
  \eta_{xxt} - \frac32 \frac{f_{xt}\eta_{xt}}{f_t} + \frac34 \frac{f_{xt}^2\eta_t}{f_t^2}
      -  3( f_x \eta_t + \eta_x f_t )  = 0  \ .  
\end{equation}
To generate local symmetries for aDP from (\ref{gensym}) we take
$s^{(1)},s^{(2)},s^{(3)}$ to be given by the 3 possible asymptotic expansions of $s$ for large $|\theta|$.
One expansion was given already in (\ref{asymp}) and the others are obtained from it by replacing
$\theta^{1/3}$ in the expansion with $\omega\theta^{1/3}$ and $\omega^2\theta^{1/3}$, 
where $\omega=\frac12 (-1+\sqrt{3}i )$ is a
cube root of $1$. Substituting these asymptotic expansions into $Q(\theta)$ and expanding in inverse powers of $\theta^{1/3}$
gives the first two nontrivial local symmetries of aDP to be
\begin{eqnarray*}
  \eta^{(1)} &=&  f_{5x} - 15f_xf_{xxx} - \frac{45}{4} f_{xx}^2 + 15 f_x^3   \ , \\ 
  \eta^{(2)} &=&  f_{7x} - 21 f_x f_{5x} - \frac{105}{2} f_{4x} f_{xx} - \frac{147}{4} f_{xxx}^2 + 126 f_x^2 f_{xxx}+ \frac{315}{2}f_xf_{xx}^2
     - 63 f_x^4 \ . 
\end{eqnarray*}
(Here $f_{4x},f_{5x},\ldots$ denote 4th, 5th etc derivatives with respect to $x$.)  
The first of these defines the potential KK flow, as expected.  

\section{Concluding Remarks}
In this paper we have introduced a BT for the aDP equation and given a superposition principle for it. We
have seen how this BT generates interesting solutions, and in particular that a single BT on the trivial solution
generates not just traveling wave solutions, but also certian ``mergers''. We have further seen that the
superposition principle of 3 BTs allows us to compute 
the hierarchy of infinitesimal symmetries, this being a further illustration of the general scheme we have proposed.

A number of matters for further work have arisen, and we briefly recap these. We have not yet translated
the solutions obtained for aDP to solutions for DP, which is of more interest because of its physical significance.
We have also indicated that the DP and aDP equations fit into a larger picture of integrable equations with a
third order Lax operator, including the KK, SK, DP and Novlkov equations, and it would be good to see the results
presented here emerge as special cases or reductions of more general results. (The fact that the generator of
infinitesimal symmetries $Q(\theta)$ for aDP is at least superficially the same as the corresponding object for
the Boussinesq equation would seem to be good evidence for this.) Finally, we have not yet found an explicit formula
for the superposition of 3 BTs of aDP, and there is good reason to believe such a thing must exist. Although we succeeded
in presenting the superposition formula in the  reasonably compact form (\ref{superpos}), this involves 4th powers of
$s_1-s_2$ and is difficult to handle. We suspect there is some underlying formalism or choice of fields
in which everything becomes much more transparent. 

\section*{Appendix: Travelling wave solutions of aDP} 

We look for travelling wave solutions of (\ref{myDP}) in the form
$$   f =   \beta x + \frac{t}{\beta} + U(\sqrt{\beta}( x-ct) )  $$
where $\beta$ is a positive constant. 
Writing  $z = \sqrt{\beta}(x-ct)$ and 
$P(z) = 1 - c\beta^{3/2} U'(z)$,  we find we need 
\begin{equation}
  P'' =  \frac34 \frac{P'^2}{P} + 3(1-P)\left( 1 -  \frac{P}{\beta^2 c} \right) \ .
\label{Pzz} 
\end{equation} 
Note that
$$ f_t(x,t)  =  \frac{P(z)}{\beta}  \ .    $$

Equation (\ref{Pzz}) can be integrated to give 
$$ P'^2  = -\frac{4}{\beta^2 c} P^3   + 12 \left(1 + \frac1{\beta^2 c} \right) P^2  + 4 d P^{3/2} + 12 P\ .   $$ 
Clearly it is advantageous to substitute $P(z)=Q(z)^2$, giving 
$$ Q'^2  =  -\frac{1}{\beta^2 c} Q^4   + 3 \left(1 + \frac1{\beta^2 c} \right) Q^2  +  d Q + 3 \ .   $$ 
For a soliton solution we require that the polynomial on the right hand side should have a double root.
Choosing $d$ so that there is a double root at $Q=1$ (corresponding to a soliton with a background value of
$f_t=\frac1{\beta})$ we obtain
\begin{equation}
  Q'^2  = -\frac{1}{\beta^2 c} \left( Q-1)^2 (Q^2 + 2 Q - 3\beta^2 c \right)\ . \label{Qeq}
\end{equation}   
If $\beta^2 c>1$, the RHS of (\ref{Qeq}) has $2$ simple real roots, in additional to the double root at $Q=1$, 
with one simple root on each side of the double root. The  corresponding solutions are 
\begin{equation}
  Q(z) = 1 +  \frac{3(\beta^2c-1)}{ 2 \pm  \sqrt{3\beta^2c+1} \cosh\left( z\sqrt{3\left(1-\frac1{\beta^2c}\right)}  \right)  }\ .
        \label{Q1}
\end{equation}   
The solution with a $+$ corresponds to the simple soliton of the Section 4 ($Q>1$ for all $z$), while 
the solution with a $-$ corresponds to the dimpled soliton ($Q$ changes sign, but $P=Q^2$ remains positive).
If $\beta^2 c<-\frac13 $, then the RHS (\ref{Qeq}) has no roots apart from the double root at $Q=1$, and is nonnegative for
all $Q$. The corresponding solutions are 
\begin{equation}
Q(z) = 1 -  \frac{3(\beta^2|c|+1)}{ 2 \pm
  \sqrt{3\beta^2|c|-1} \sinh\left( z\sqrt{3\left(1+\frac1{\beta^2|c|}\right)}  \right)  }\ .   \label{Q2}
\end{equation}  
These correspond to the singular solitons of Section 4.  
It is straightforward to verify that (\ref{Q1}) and (\ref{Q2}) are the only solitary wave solutions. 

\bibliographystyle{acm}
\bibliography{cp}
\end{document}